\documentclass{article}
\begin{document}
\title{Lessons from numerical analysis}
\author{A. Rivero\thanks{Dep. Informatica, Univ Carlos III Madrid, Spain}}
\maketitle
\begin{abstract}
It is reviewed how Renormalization Group, species doubling and CKM mixing 
are known to appear in numerical analysis: Butcher group, parasitic 
solutions and higher order methods  
\end{abstract}

\section{Butcher Group}

In the late sixties Butcher \cite{butcher} found a composition rule
for Runge Kutta methods by using a multiplication of rooted
trees. Recently the same multiplication has been found in the
perturbative renormalization of Feynman diagrams (Kastler) and
in the group of diffeomorphisms of a manifold (Connes and Moscovici).
The recent findings were done under the cover of Hopf algebras, but
C. Brouder\cite{brouder} pointed the similarity with the previous work 
of Butcher.

Butcher group, as it is, has not enough capacity to hold the quantity
of renormalized parameters in a QFT theory, so it is usually extended to
decorated trees. Again, this is not unusual, as the treatment of
methods beyond Runge Kutta (numerical geometrical integration methods,
for instance) also uses labeled nodes to classify them.

An up-to-date review of the use of this structure in Numerical Analysis,
including some kinds of decorated trees, will be found in the
monograph \cite{hairer} of the Geneva team. Unlabeled trees are well
covered in the books of Butcher and Hairer and Wanner.   
 
Composition of Runge Kutta methods implies scale change in a way very similar
to Wilson's view of the renormalization group. A RK method (or two different
ones) can be applied two times, from $y(x)$ to $y(x+h)$ and then, taking
$y(x+h)$ as starting value, to $y(x+2h)$. But the composition is again a
RK method, from $y(x)$ to $y(x+h)$, with a more complicated description, in
the same way that a block-spin composition gives as a new lattice Hamiltonian
with a more complicated action.

It is possible to build a method, with an infinite (continuous) set of
RK parameters, giving as result the exact solution of the integration
problem. This is called the Picard method, and its associated
B-series can be seen to have special properties preserving its form
under scaling.

The undecorated Butcher group is powerful enough to control the change
of scales in a structure pertinent to quantization, the tangent groupoid.
Continuous functions over this groupoid are known to be determined
by Weyl quantization, and the points of the groupoid are parametrized
by a variable that jumps from a role of cutoff to a role of Plank
constant. This should happen by a spontaneous apparition of scale associated
to the choosing of renormalization point, but the details are of the
action of Butcher Group inside the groupoid are still to be studied.

\section{Doubling}

It was noticed by Rothe \cite[p.48 ff]{rothe} that the multiplicity of 
solutions
in multi-steps methods, in the particular
case of the symmetric
derivative method, corresponds to the fermion doubling of lattice QCD.

Numerical methods rule out the doubled solution by imposing a continuity
in the solution, which it turn imposes to the parasitic solution a 
coefficient of 
order $h^2$. This is enough for the method to work, although it remains
weakly unstable.

In field theory, for fermions, the extra solution can hide under the cover
of a different representation of gamma matrices, and it can not be 
eliminated. So $d$-dimensional field theory will produce $2^d$ fermionic
degrees of freedom when discretized. 

In lattice QFT, $d=4$, the doubling can be reduced to four species by using
Susskind formulation. This is just the discrete version of Dirac-Kahler 
equation,  a old recipe from K\"aler where
differential forms are used to implement a Dirac equation containing four
 degenerate
species of fermions. It is unknown if degeneracy can be broken in order to
obtain the four fundamental fermions: there is a huge difference of masses
and charges, specially the confinement of two fermions. In \cite{rivero} it
is speculated that the 
quark sector $u,d$ should
be seen as differential forms corresponding to angular coordinates, while the
leptonic sector $e,\nu$ should be the radial part of a volume form 
$d\theta \wedge d\phi \wedge dr \wedge dt$. But no further work has been done
in this direction.

\section{Cabbibo angles}

If we expand a function in powers of the cutoff,

$$f(x+nh)= f(x) + n f'(x) h + {n^2 \over 2} f''(x) h^2 + ... $$

we can see the basis of multi-step methods: by choosing adequate combinations
of $f(x),f(x-h),f(x-2h),...$ it is possible to approach $f'(x)$ at higher orders
of the step size $h$. 
For instance, if we work with $f(x), f(x-h), f(x+h)$, the symmetric derivative
approaches the continuous one with order $h^2$. Besides chiral preservation,
this is a
 motivation to work
with symmetric derivatives. Forward or backward derivatives have only
order $h$.

Now, for higher orders, the symmetric solution, or the combination
of symmetric derivatives, is not optimal. So an optimal fixing of the
combination must be found if we need high precision or, also, if we
need to approach further derivatives $f',f'',...$

Suppose you want to calculate
$f''(x)$ by nesting some discrete derivations 
$(f(x+a_i+\beta_i h)-f(x+a_i))/\beta_i h$. Then the optimal combination
will correspond to some adimensional relationships between parameters. If 
derivations are formulated as an NCG in discrete space, such relationships
can be made to appear
as a CKM matrix.

\end{document}